\documentclass[12pt,epsf]{article}
\input epsf
\begin{document}
\title{Problem of Light Scalar Mesons\thanks{Talk given on May 28, 2004 at QUARK-2004,
 May 24-30, 2004, Pushkinskie Gory, Pskovskaya region, Russia}}
\author{N.N. Achasov
\thanks{Email address: achasov@math.nsc.ru}\\[1cm]
Laboratory of Theoretical Physics,\\
 Sobolev Institute for Mathematics,\\
 Academician Koptiug Prospekt, 4,
  Novosibirsk, 630090, Russia}
  \date{\today}
 \maketitle
\begin{abstract}
The following topics are considered.
\begin{enumerate}
 \item Confinement, chiral dynamics and light scalar
  mesons.
 \item  $\phi$-meson radiative decays about  nature of light scalar
    resonances.
 \end{enumerate}
Arguments in favor of the four-quark model of the $a_0(980)$ and
$f_0(980)$ mesons are given.
\end{abstract}

 \section{Kategorischer Imperativ}

 To discuss actually the nature of the  putative nonet of the light scalar mesons:
  the red putative $f_0(600)$ (or $\sigma (600)$) and $\kappa
 (700-900)$
mesons and the red well established $f_0(980)$ and $a_0(980)$
mesons,
  one should explain not only their
mass spectrum, particularly the mass degeneracy of the $f_0(980)$
and $a_0(980)$ states, but answer the next real challenges.
 \begin{enumerate}
\item  The copious  $\phi\to\gamma f_0(980)$ decay
and  especially  the copious $\phi\to\gamma a_0(980)$ decay, which
looks as the decay plainly forbidden by the Okubo-Zweig-Izuka
(OZI) rule in the quark-antiquark model $a_0(980)=(u\bar u - d\bar
d)/\sqrt{2}$.
\item  Absence of  $J/\psi\to a_0(980)\rho$ and
$J/\psi\to f_0(980)\omega$   with copious $J/\psi\to
a_2(1320)\rho$, $J/\psi\to
 f_2(1270)\omega$, $J/\psi\to f_2^\prime (1525)\phi$
 if  $a_0(980)$ and $f_0(980)$ are
 $P$-wave states of $q\bar q$ like
$a_2(1320)$ and $f_2(1270)$ respectively.
\item  Absence of  $J/\psi\to\gamma f_0(980)$   with copious
$J/\psi\to\gamma f_2(1270)$ and $J/\psi\to\gamma f_2^\prime
(1525)\phi$  if $f_0(980)$ is
 $P$-wave state of $q\bar q$ like
 $f_2(1270)$ or
$f_2^\prime(1525)$.
\item  Suppression of $a_0(980)\to\gamma\gamma$ and $f_0(980)\to\gamma\gamma$  with copious
 $a_2(1320)\to\gamma\gamma$,
$f_2(1270)\to\gamma\gamma$, $f_2^\prime(1525)\to\gamma\gamma$ if
$a_0(980)$ and
 $f_0(980)$ are
 $P$-wave state of $q\bar q$ like
$a_2(1320)$ and
 $f_2(1270)$ respectively.
\end{enumerate}

Unfortunately, I have opportunity to dwell on the $\phi$ radiative
decays only in this talk.

\section{Place in QCD, Chiral Limit,Confinement, \boldmath{$\sigma$}-Models, History, Chiral Shielding,
Troubles and Expectancies}

 Study of the nature of light
scalar resonances has become a central problem of non-perturbative
of QCD. The point is that the elucidation of their nature is
important for understanding both  the confinement physics and the
chiral symmetry realization way in the low energy region, i.e.,
the main consequences of QCD in the hadron world.
\begin{equation}
 L=-(1/2)Tr\left (G_{\mu\nu}G^{\mu\nu}\right )+\bar
q(\imath\hat{D}-M)q,\nonumber
\end{equation}
\begin{displaymath}
 M=\left(\begin{array}{ccc}
 m_u & 0& 0 \\
 0 & m_d & 0 \\
 0 & 0 & m_s \\
 \end{array} \right),\quad
 q^\alpha\equiv q^\alpha(x)=
 \left(\begin{array}{c}
 u^\alpha(x) \\
d^\alpha(x) \\ s^\alpha(x)\\
 \end{array} \right)
 \end{displaymath}
  $M$ mixes Left and Right Spaces
 $q_L(x)=(1/2)(1+\gamma_5)q(x)$ and $q_R(x)=(1/2)(1-\gamma_5)q(x)$.
 But in  chiral limit  $M_{ij}\to 0$  these spaces
 separate realizing   $U_L(3)\times U_R(3)$ flavour symmetry,
 $q^\prime_{L,R}(x)=\mbox{U}_{L,R}\cdot q_{L,R}(x)$,   which, however, is broken by the gluonic anomaly
up to $U_{vec}(1)\times SU_L(3)\times SU_R(3)$. As Experiment
suggests, Confinement forms colourless observable hadronic fields
and spontaneous breaking of chiral symmetry with massless
pseudoscalar fields. There are two possible scenarios for OCD at
low energy.
\begin{enumerate}
\item Non-linear $\sigma$-model.
\begin{eqnarray}
&& L=\left(F_\pi^2/2\right)Tr\left(\partial_\mu V(x)\partial^\mu
V^+(x)\right) +...\,,\nonumber\\ &&  V(x)=\exp\left
\{2\imath\phi(x)/F_\pi\right\},\quad
V^\prime(x)=\mbox{U}_LV(x)\mbox{U}_R\,.\nonumber
\end{eqnarray}
\item Linear $\sigma$-model.
\begin{eqnarray}
&& L=\left(1/2\right)Tr\left(\partial_\mu \mbox{V}(x)\partial^\mu
\mbox{V}^+(x)\right) -
W\left(\mbox{V}(x)\mbox{V}^+(x)\right)\,,\nonumber\\ &&
\mbox{V}(x)=\left(\sigma(x)+\imath\pi(x)\right),\quad
\mbox{V}^\prime(x)=\mbox{U}_L\mbox{V}(x)\mbox{U}_R\,.\nonumber
\end{eqnarray}
\end{enumerate}

 The experimental
nonet of the light scalar mesons, the  putative $f_0(600)$ (or
$\sigma (600)$) and $\kappa (700-900)$ mesons and the well
established $f_0(980)$ and $a_0(980)$ mesons as if suggests the
$U_L(3)\times U_R(3)$ linear $\sigma$-model.

 Hunting   the light $\sigma$ and $\kappa$
mesons had begun in the sixties already and a preliminary
information on the light scalar mesons in Particle Data Group
(PDG) Reviews had appeared at that time. But long-standing
unsuccessful attempts to prove their existence in a  conclusive
way entailed general disappointment and an information on these
states disappeared from PDG Reviews. One of principal reasons
against the $\sigma$ and $\kappa$ mesons was the fact that both
$\pi\pi$ and $\pi\kappa$  scattering phase shifts  do not pass
over $90^0$ at putative resonance masses.

 Situation
changes when we showed (1994)  that in the linear $\sigma$-model
\begin{eqnarray}
&&  L=\left(1/2\right)\left
[(\partial_\mu\sigma)^2+(\partial_\mu\vec{\pi})^2\right ]+
\left(\mu^2/2\right)\left [(\sigma)^2+(\vec{\pi})^2\right ]
\nonumber\\ && - \left(\lambda/4\right)\left
[(\sigma)^2+(\vec{\pi})^2\right ]^2 \nonumber
\end{eqnarray}
 there is a negative background
phase which hides the $\sigma$ meson. It has been made clear that
shielding wide lightest scalar mesons in chiral dynamics is very
natural.

 This idea was picked up and triggered
new wave of theoretical and experimental searches for the $\sigma$
and $\kappa$ mesons.

We considered the simplest Dyson equation for the $\pi\pi$
scattering amplitude with  real intermediate $\pi\pi$ states only
(1994).

\begin{eqnarray}
 && T_0^{0(tree)}=\frac{m_\pi^2-m_\sigma^2}{16\pi F^2_\pi}\left
[5-3\frac{m_\sigma^2-m_\pi^2}{m_\sigma^2-s}
-2\frac{m_\sigma^2-m_\pi^2}{s-4m_\pi^2}\ln\left
(1+\frac{s-4m^2_\pi}{m_\sigma^2}\right )\right ],\nonumber\\[24pt]
&&
 T^0_0=\frac{T_0^{0(tree)}}{1-\imath\rho_{\pi\pi}T_0^{0(tree)}}=
 {\frac{e^{2\imath\left(\delta_{bg}+\delta_{res}\right )}-1}{2\imath\rho_{\pi\pi}}
=\frac{1}{\rho_{\pi\pi}}\left [\left
 (\frac{e^{2\imath\delta_{bg}}-1}{2\imath}\right )+
 e^{2\imath\delta_{bg}}T_{res}\right ]},\nonumber\\[24pt]
&& T_{res}=\frac{\sqrt{s}\Gamma_{res}(s)}{M^2_{res} - s +
\Re(\tilde \Pi_{res}(M^2_{res}))-\tilde
\Pi_{res}(s)}=\frac{e^{2\imath\delta_{res}}-1}{2\imath\rho_{\pi\pi}}\,.\nonumber
\end{eqnarray}

In theory the  principal problem is  impossibility to use the
linear $\sigma$-model in the tree level approximation inserting
widths into $\sigma$ meson propagators because such an approach
breaks the both  unitarity and  Adler self-consistency conditions,
as we showed (1994). Strictly speaking, the comparison with the
experiment  requires the non-perturbative calculation of the
process amplitudes. Nevertheless, now there are the possibilities
to estimate odds of the $U_L(3)\times U_R(3)$ linear
$\sigma$-model to underlie physics of light scalar mesons  in
phenomenology. Really, even now there is a huge body of
information about the $S$- waves
 of different two-particle
pseudoscalar states and what is more  the relevant information go
to press almost continuously from BES, BNL, CERN, CESR,
DA$\Phi$NE, FNAL, KEK, SLAC and others.  As for theory, we know
quite a lot about the  scenario under discussion: the nine scalar
mesons, the putative chiral shielding of the $\sigma (600)$ and
$\kappa( 700-900)$ mesons, the  unitarity and  Adler
self-consistency conditions. In addition,  there is the light
scalar meson treatment  motivated by field theory. The foundations
of this approach were formulated in our papers (1979-1984).

\section{Four-quark Model}

 The nontrivial nature of the
well established light scalar resonances $f_0(980)$ and $a_0(980)$
is no longer denied practically anybody. In particular, there
exist numerous evidences in favour of the $q^2\bar q^2$ structure
of these states. As for the nonet as a whole, even a dope's look
at PDG Review gives an idea of the four-quark structure of the
light scalar meson nonet, $\sigma (600)$, $\kappa (700-900)$,
$f_0(980)$, and $a_0(980)$, inverted in comparison with the
classical $P$-wave $q\bar q$ tensor meson nonet, $f_2(1270)$,
{$a_2(1320)$, $K_2^\ast(1420)$, $\phi_2^\prime (1525)$. Really,
while the scalar nonet  cannot be treated as the $P$-wave $q\bar
q$ in the naive  quark model, it can be easy understood as the
$q^2\bar q^2$ nonet, where $\sigma (600)$ has  no strange quarks,
$\kappa (700-900)$ has the $s$ quark, $f_0(980)$ and $a_0(980)$
have the $s\bar s$ pair,  R.L. Jaffe, J. Schechter et al..

 The  scalar mesons $a_0(980)$ and
$f_0(980)$, discovered more than thirty years ago, became the hard
problem for the naive $q\bar q$ model from the outset. Really, on
the one hand the almost exact degeneration of the masses of the
isovector $a_0(980)$ and isoscalar $f_0(980)$ states revealed
seemingly the structure similar to the structure of the vector
$\rho$ and $\omega$ mesons,  and on the other hand the strong
coupling of $f_0(980)$ with the $K\bar K$ channel as if suggested
a considerable part of the strange pair $s\bar s$ in the wave
function of $f_0(980)$.

 In 1977
R.L. Jaffe  noted that in the MIT bag model, which incorporates
confinement phenomenologically, there are light four-quark scalar
states. He suggested also that $a_0(980)$ and $f_0(980)$ might be
these states with symbolic structures
\begin{eqnarray}
&& a^0_0(980)=(us\bar u\bar s - ds\bar d\bar s)/\sqrt{2} \quad
\mbox{and}
 f_0(980)=(us\bar u\bar s + ds\bar d\bar
s)/\sqrt{2}.\nonumber
\end{eqnarray}
 From that time $a_0(980)$ and $f_0(980)$
resonances came into  beloved children of the light quark
spectroscopy.

\section{Radiative Decays of \boldmath{$\phi$}-Meson }

 Ten years later we showed that the study of the radiative decays
$\phi\to\gamma a_0\to\gamma\pi\eta$ and $\phi\to\gamma f_0\to
\gamma\pi\pi$ can shed light on the problem of $a_0(980)$ and
$f_0(980)$ mesons. Over the next ten years before experiments
(1998) the question was considered from different points of view.
Now these decays have been studied not only theoretically but also
experimentally. The first measurements have been reported by the
SND  and CMD-2 Collaborations which obtain the following branching
ratios
\begin{eqnarray}
&& BR(\phi\to\gamma\pi^0\eta)= (0.88\pm 0.14\pm 0.09)\cdot10^{-4}\
\ \mbox{SND},\nonumber\\ && BR(\phi\to\gamma\pi^0\pi^0)= (1.221\pm
0.098\pm 0.061)\cdot10^{-4}\ \ \mbox{SND},\nonumber\\
 &&
BR(\phi\to\gamma\pi^0\eta)=(0.9\pm 0.24\pm 0.1)\cdot10^{-4}\ \
\mbox{CMD-2},\nonumber\\
 &&
BR(\phi\to\gamma\pi^0\pi^0)=(0.92\pm0.08\pm0.06)\cdot10^{-4}\ \
\mbox{CMD-2}.\nonumber
\end{eqnarray}}

 More recently the KLOE Collaboration has measured
\begin{eqnarray}
&&
BR(\phi\to\gamma\pi^0\eta(\to\gamma\gamma))=(0.851\pm0.051\pm0.057)\cdot
10^{-4}\nonumber\\
 &&
BR(\phi\to\gamma\pi^0\eta(\to\pi^+\pi^-\pi^0)
)=(0.796\pm0.060\pm0.040)\cdot10^{-4}\nonumber\\
&&
BR(\phi\to\gamma\pi^0\pi^0)=
(1.09\pm0.03\pm0.05)\cdot10^{-4}\nonumber
\end{eqnarray}
in agreement with the Novosibirsk data but with a considerably
smaller error.

Note that $a_0(980)$  is produced in the radiative $\phi$ meson
decay
 as intensively as $\eta '(958)$  containing $\approx 66\% $ of $s\bar s$,
  responsible for $\phi\approx s\bar s\to\gamma s\bar s\to\gamma \eta '(958)$.
 It is a clear qualitative argument
for the presence of the $s\bar s$ pair in the isovector $a_0(980)$
state, i.e., for its four-quark nature.

\subsection{\boldmath{$K^+K^-$}-Loop Model}

 When basing the experimental investigations, we suggested
one-loop model $\phi\to K^+K^-\to\gamma a_0(980)\, (\mbox{ or}\,
f_0(980))$. This model is used in the data treatment and is
ratified by experiment.

 Below we argue on gauge invariance
grounds that the present data give the conclusive arguments in
favor of the $K^+K^-$-loop transition as the principal mechanism
of $a_0(980)$ and $f_0(980)$ meson production in the $\phi$
radiative decays. This enables to conclude
 that production of the lightest scalar mesons $a_0(980)$ and
 $f_0(980)$ in these decays
  is caused by the four-quark transitions, resulting in strong restrictions on the
 large $N_C$ expansions of the decay amplitudes. The analysis
 shows that these constraints give new evidences in favor
 of the four-quark nature of $a_0(980)$ and $f_0(980)$
 mesons.

\subsection{Spectra and Gauge Invariance}

 To describe the experimental spectra
\begin{eqnarray}
&& \frac{dBR(\phi\to\gamma R\to\gamma ab\,,\,
m)}{dm}=\frac{2}{\pi}\frac{m^2\Gamma(\phi\to\gamma
R\,,\,m)\Gamma(R\to ab\,,\,m)}{\Gamma_\phi|D_R(m)|^2}\nonumber\\
&& = \frac{4|g_R(m)|^2\omega (m) p_{ab}(m)}{\Gamma_\phi\,
3(4\pi)^3m_{\phi}^2}\left |\frac{g_{Rab}}{D_R(m)}\right |^2,\quad
R=a_0,\, f_0,\ ab=\pi^0\eta,\ \pi^0\pi^0,\nonumber
\end{eqnarray}
the function $|g_R(m)|^2$ should be smooth (almost constant) in
the range $m\leq 0.99$ GeV. But the problem issues from gauge
invariance which requires that
\begin{eqnarray}
 && A\left [\phi(p)\to\gamma (k) R(q)\right
]\nonumber\\[3pt] &&= G_R(m)\left [p_\mu e_\nu(\phi) - p_\nu
e_\mu(\phi)\right]\left [k_\mu e_\nu(\gamma) - k_\nu
e_\mu(\gamma)\right].\nonumber
\end{eqnarray}
Consequently, the function
\begin{eqnarray}
g_R(m)= - 2(pk)G_R(m) = - 2\omega (m) m_\phi G_R(m)\nonumber
\end{eqnarray}
is proportional to the photon energy
$\omega(m)=(m_{\phi}^2-m^2)/2m_{\phi}$ (at least!) in the soft
photon region.

Stopping the function $(\omega (m))^2$ at $\omega
(990\,\mbox{MeV})=29$ MeV with the help of the form-factor
$1/\left [1+(R\omega (m))^2\right ]$ requires $R\approx 100$
GeV$^{-1}$. It seems to be incredible to explain such a huge
radius in hadron physics.
 Based on rather great
 $R\approx 10$ GeV$^{-1}$, one can
 obtain an effective maximum of the mass spectrum only near 900
MeV.

 To exemplify this trouble let us consider the contribution of
the isolated $R$ resonance: $g_R(m)=-2\omega (m) m_\phi G_R\left
(m_R\right )$. Let also the mass and the width of the $R$
resonance equal 980 MeV and 60 MeV, then
$S_R(920\,\mbox{MeV}):S_R(950\,\mbox{MeV}):S_R(970\,\mbox{MeV}):S_R(980\,\mbox{MeV})
=3:2.7:1.8:1$.

 So stopping the $g_R(m)$ function is the crucial
point in understanding  the mechanism  of the production of
$a_0(980)$ and $f_0(980)$ resonances in the $\phi$ radiative
decays.

\begin{figure}
\centerline{\epsfxsize=14cm \epsfysize=8.5cm \epsfbox{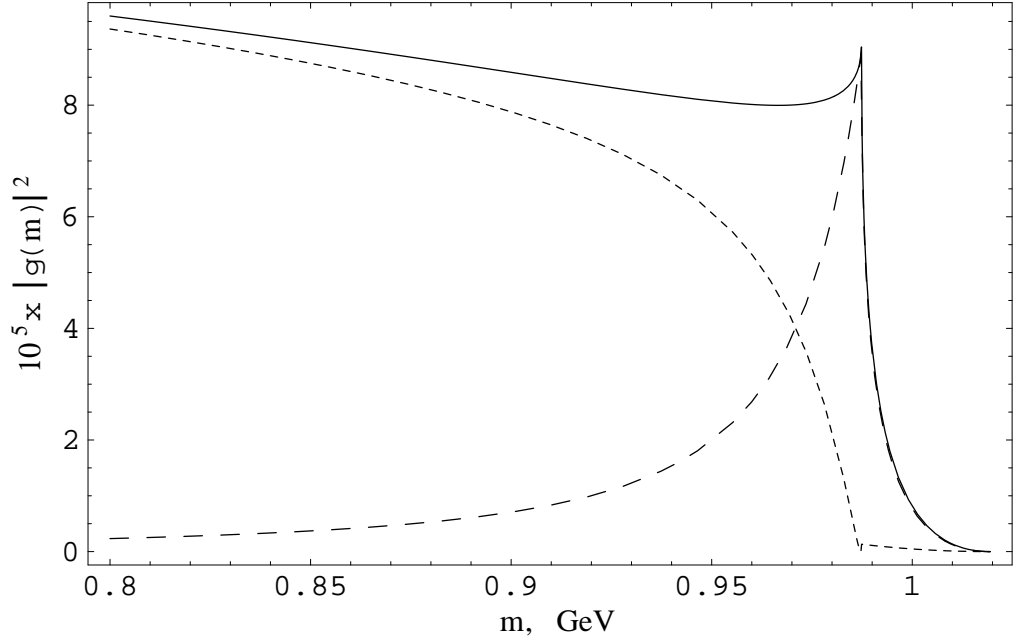}}
\caption{
 The universal in the $K^+K^-$ loop model function
$|g(m)|^2=\left |g_R(m)/g_{RK^+K^-}\right |^2$ is drawn with the
solid line. The contribution of the
 imaginary part is drawn with the dashed line. The contribution of the real part
   is drawn with the dotted line.
   }
 \label{g}
\end{figure}

 The $K^+K^-$-loop model $\phi\to K^+K^-\to\gamma
R$ solves this problem in the elegant way:   fine threshold
phenomenon is discovered, see Fig. \ref{g}.

\begin{figure}
\centerline{\epsfxsize=12cm \epsfysize=8cm \epsfbox{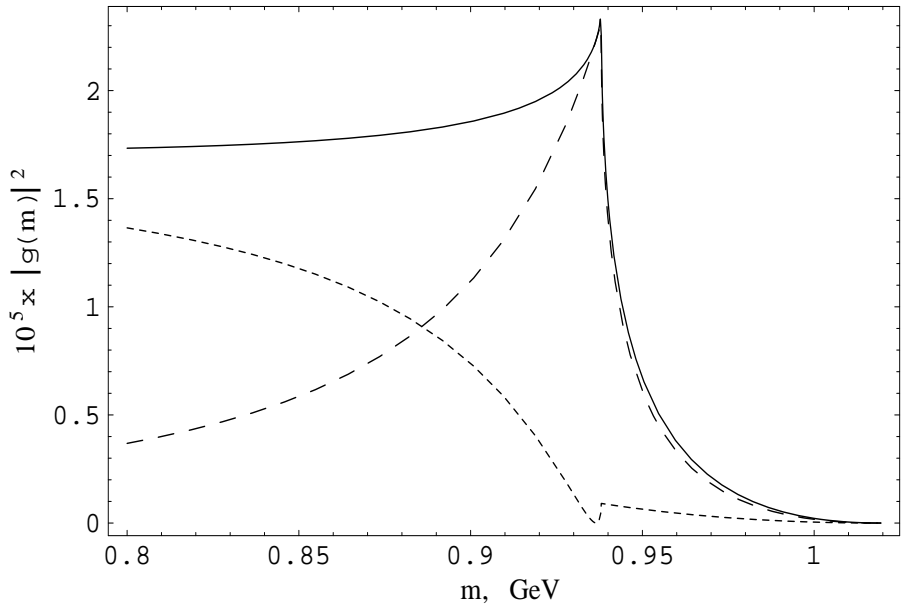}}
 \caption{
  The function $|g(m)|^2$ for  $m_{K^+}=469$ MeV is drawn
  with the solid line. The contribution of the imaginary part is drawn with the dashed line.
  The contribution of the real part is drawn with the dotted line.
 }
   \label{gg}
\end{figure}

\begin{figure}
\centerline{\epsfxsize=12cm \epsfysize=8cm \epsfbox{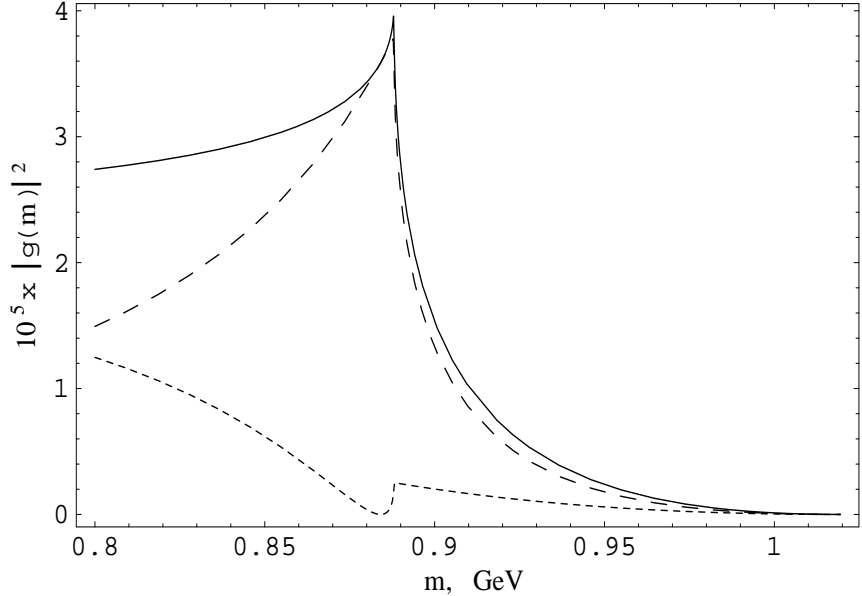}}
\caption{
     The function $|g(m)|^2$ for  $m_{K^+}=444$ MeV is drawn
     with the solid line. The contribution of the imaginary part is drawn with the dashed line.
     The contribution of the real part is drawn with the dotted line.
} \label{ggg}
\end{figure}

To demonstrate the threshold character of this effect we present
 Fig. \ref{gg} and Fig. \ref{ggg} in which the function $|g(m)|^2$
is shown in the case of $K^+$ meson mass is 25 MeV and 50 MeV less
than in reality.

 As  seen from   Figs. \ref{gg} and \ref{ggg}, the function
$|g(m)|^2$ is suppressed by the $(\omega (m))^2$ law in the region
950-1020 MeV and 900-1020 Mev respectively. In the mass spectrum
this suppression is increased by one more power of $\omega (m)$,
so that we cannot see the resonance in the region 980-995 MeV .
The  spectrum maximum  is effectively shifted to the region
935-950 MeV and 880-900 MeV respectively.

 In truth this means that
$a_0(980)$ and $f_0(980)$ resonances are seen in the radiative
decays of $\phi$ meson owing to the $K^+K^-$ intermediate state,
otherwise the maxima in the spectra would be shifted to 900 MeV.

So the mechanism of production of $a_0(980)$ and $f_0(980)$ mesons
in the $\phi$ radiative decays is established.

\subsection{Four-quark Transition}

 Both real and imaginary parts of the $\phi\to\gamma
R$  amplitude are caused by the $K^+K^-$ intermediate state. The
imaginary part is caused by the real $K^+K^-$ intermediate state
while the real part is caused by the virtual compact $K^+K^-$
intermediate state, i.e., we are dealing here with  the four-quark
transition.
 \footnote{ It will be recalled that the imaginary part
of every hadronic amplitude describes a multi-quark transition.}
 Needless to say, radiative four-quark
transitions can happen between two $q\bar q$ states as well as
between $q\bar q$ and $q^2\bar q^2$ states but their intensities
depend strongly on a type of the transitions.

\subsection{Four-quark Transition, OZI rule, and large \boldmath{$N_C$} expansion}
 A radiative four-quark transition between two $q\bar
q$ states requires creation and annihilation of an additional
$q\bar q$ pair, i.e., such a transition is forbidden according to
the Okuba-Zweig-Izuka (OZI) rule, while a radiative four-quark
transition between $q\bar q$ and $q^2\bar q^2$ states requires
only creation of an additional $q\bar q$ pair, i.e., such a
transition is allowed according to the OZI rule.

 Let us
consider this problem from the large $N_C$ expansion standpoint,
using the G.'t Hooft rules : $g^2_sN_C\to const$ at $N_C\to\infty$
and a gluon is equivalent to a quark-antiquark pair ( $A^i_j\sim
q^i\bar q_j$ ).

  The point is that the large $N_C$ expansion is the
most clear heuristic understanding of the OZI rule because the OZI
forbidden branching ratio as a rule is suppressed by the factor
$N_C^2=9$, but not $N_C=3$, in comparison with the OZI allowed
decay. In our case the results of the analysis are even more
interesting.

\subsection{Summary on \boldmath{$\phi$} radiative decays from the
large \boldmath{$N_C$} expansion standpoint }

 The analysis shows   that the $\phi\to K^+K^-\to\gamma a_0$
decay intensity in the two-quark model  and the $\phi\to
K^+K^-\to\gamma f_0$ decay intensity in the $f_0=(u\bar u + d\bar
d)/\sqrt 2$ model are suppressed by the factor $1/N_c^3$ in
comparison with the ones in the four-quark model, but in the
$f_0=s\bar s$ model the $\phi\to K^+K^-\to\gamma f_0$ decay
intensity is suppressed only by the factor $1/N_c$ in comparison
with the one in the four-quark model. In this model, in addition
to the serious trouble with the $a_0-f_0$ mass degeneration,  the
$\left (N_C\right )^0$ order transition without creation of an
additional $q\bar q$ pair $\phi\approx s\bar s\to\gamma s\bar
s\to\gamma f_0(980)$ ( similar to the principal mechanism of the
$\phi\approx s\bar s\to\gamma s\bar s\to\gamma\eta ' (958)$ decay)
is  bound to have  a small weight in the large $N_C$ expansion of
the $\phi\approx s\bar s\to\gamma f_0(980)$ amplitude, because
this term does not contain the $K^+K^-$ intermediate state, which
emerges only in the next to leading term of the $1/N_C$ order,
i.e., in the OZI forbidden transition. So, in this model the
$\phi\approx s\bar s\to\gamma f_0(980)$ amplitude has the $1/N_C$
order  like the $\phi\approx s\bar s\to\gamma\pi^0$ one. Emphasize
that the mechanism without creation and annihilation of the
additional $u\bar u$ pair cannot explain the $f_0(980)$ spectrum
because it does not contain the $K^+K^-$ intermediate state!

\subsection{Conclusion of
\boldmath{$\phi$}-radiative decays}

 So, the fine threshold phenomenon is discovered, which is to say
that the $K^+K^-$ loop mechanism of
 the $a_0(980)$ and $f_0(980)$ scalar meson production in the
$\phi$ radiative decays is established at a physical level of
proof. The case is rarest in hadron physics. This production
mechanism is the four-quark transition what constrains the large
$N_C$ expansion of the $\phi\to\gamma a_0(980)$ and $\phi\to\gamma
f_0(980)$ amplitudes and gives the new strong ( if not crucial)
evidences in favor of the four-quark nature of $a_0(980)$ and
$f_0(980)$ mesons.
\begin{center}
{\bf\large Acknowledgments}
\end{center}
I thank Organizers of QUARKS-2004 very much for the kind
invitation, the generous hospitality, and the financial support.

This work was supported in part by the RFBR Grant No. 02-02-16061
and the Presidential Grant No. 2339.2003.2 for support of Leading
Scientific Schools.\\[2pc] {\it\large NB} {\bf\large References}
contain only author's articles on basis of which the present talk
has been written. Detailed references to other authors are in
these articles.

\end{document}